\documentclass[11pt]{article}
\usepackage{moriond,epsfig}

\def\Journal#1#2#3#4{{#1} {\bf #2}, #3 (#4)}

\def\NPB{{\em Nucl. Phys.} B}
\def\PLB{{\em Phys. Lett.}  B}

\def\PRD{{\em Phys. Rev.} D}
\def\ZPC{{\em Z. Phys.} C}
\def\EPJC{{\em Eur.Phys.J.} C}
\def\SPJ{\em Sov.Phys JETP}
\def\SJNP{\em Sov.J.Nucl.Phys.}
\def\CPC{\em Comp.Phys.Comm.}

\def\be{\begin{equation}}
\def\ee{\end{equation}}
\def\bea{\begin{eqnarray}}
\def\eea{\end{eqnarray}}

\begin{document}
\vspace*{4cm}
\title{REAL AND VIRTUAL PHOTON STRUCTURE}

\author{ L. J\"ONSSON}

\address{Department of Physics, Lund University, Box 118,\\
221 00 Lund, Sweden}

\maketitle\abstracts{The structure of real and virtual photons has been 
studied in electron-proton scattering processes producing di-jet events 
at HERA by the H1 and ZEUS collaborations. Data have been compared to 
next-to-leading order QCD calculations and to the predictions of Monte Carlo 
generators based on the DGLAP and CCFM formalisms for describing the 
parton dynamics.}

\section{Introduction}
Electron-proton scattering proceeds via the exchange of a virtual photon.
For low values of the photon virtuality i.e. $Q^2$ close to zero we
are in the region of photoproduction and as the virtuality increases 
above $Q^2 \sim$ 1 GeV$^2$ we enter the region of deep inelastic scattering
(DIS).
Electron-proton scattering events, which contain jets of high transverse 
momenta can be used to 
probe the partonic structure of either the proton or the exchanged photon, 
depending on the kinematic situation. As long as the transverse momentum 
of the parton propagator entering the hard scattering process is larger
than the vituality of the photon we will probe the constituents of the 
photon. This is obviously always true in photoproduction of high-p$_t$ 
jets but may also be 
the case in DIS. Thus, comparisons with predictions of NLO perturbative 
QCD calculations and of Monte Carlo models will provide information about 
the structure of real as well as virtual photons.

We can distinguish between two classes of events where in one class
the photon interacts as a pointlike particle (so called direct processes)
and in the other the photon interacts via its partonic constituents 
(resolved photon processes). In the BFKL \cite{bfkl} description of the parton
dynamics both these situations are taken into account in a natural way due 
to the arbitrary ordering of transverse momenta in the initial state parton 
ladder.

Experimentally the scaled photon energy, $x_\gamma$, is sensitive to the 
momenta of the incoming partons and can be used to separate direct photon
processes from resolved ones. $x_\gamma$-spectra from H1 and ZEUS show that 
a cut at around 0.7-0.8 gives a good separation between these classes of events.

\section{The virtual photon structure}

The virtual photon structure has been investigated by the H1 and 
ZEUS-collaborations in terms of triple differential di-jet cross sections,
$d^3\sigma / dQ^2 dx_\gamma dy$.

\begin{figure}
\vskip 0.1cm
\begin{center}
\epsfig{figure=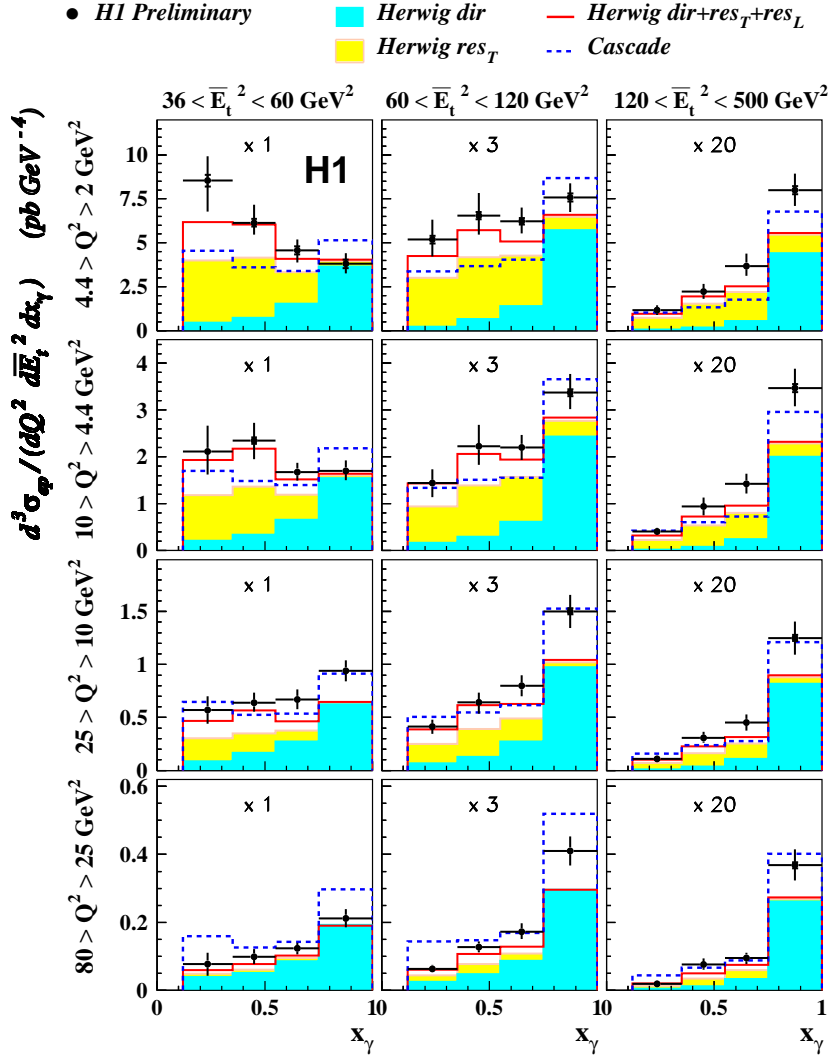,height=6in}
\end{center}
\caption{Tripple differential cross section as a fuction of $x_\gamma$ 
in bins of $Q^2$ and $E_t^2$. Data are compared to the predictions of
the HERWIG and CASCADE Monte Carlo generators.
\label{tridiff}}
\end{figure}

Fig.\ref{tridiff} shows results, obtained by the H1-experiment, on 
$d\sigma / dx_\gamma$ in bins of $Q^2$ and the transverse jet energy squared, 
$E_t^2$. The data are compared to predicitons from the HERWIG Monte Carlo
program \cite{herwig} assuming direct processes only (dark shaded histogram), direct 
plus resolved processes (light shaded histogram) \cite{jjk}, and with the 
contribution of longitudinally polarised photons added (full line 
histogram) \cite{chyla}. In addition the prediction of the CASCADE Monte Carlo 
program \cite{cascade}
is given (broken line histogram). CASCADE is based on the CCFM 
formalism \cite{ccfm}, 
which in a consistent way implements the DGLAP \cite{dglap} and BFKL evolution
equations. The figure clearly shows that direct processes alone do not
describe the data although the agreement is improved as $Q^2$ increases
with respect to $E_t^2$. The addition of resolved processes improves the
situation over the whole kinematic range and even more so if the contribution
from longitudinally polarised photons is taken into consideration.
CASCADE with less degrees of freedom offers a description of the data which 
is of a similar quality as HERWIG including all photon contributions.

\section{The real photon structure}
 
The structure of real photons has been investigated by measuring
the di-jet cross section in photoproduction at high transverse energies 
where a comparison with next-to-leading order (NLO) calculations is expected to be 
valid. In the measured kinematic domain non-perturbative effects like multiple 
interactions and hadronisation have been 
found to be small. The NLO calculation is based on the subtraction method 
\cite{subtr} in order to handle divergencies
from collinear and infra-red emissions. Different parametrisations 
of the parton densities of the proton (CTEQ5M1 \cite{cteq}, MRST99 \cite{mrst}) 
and the photon (GRV-HO \cite{grv}, AFG-HO \cite{afg}) were used to 
investigate the dependence of the 
NLO cross section to the parton density functions.

The H1 collaboration observes good agreement, over the full
kinematic region investigated, between data and
the NLO calculations with little dependence on the photon structure functions.

The ZEUS collaboration, on the other hand, 
finds that none of the photon structure functions used
in the NLO calculations is able to give a fully satisfactory description of 
the data. Especially in kinematic regions where the photon structure function 
is expected to have a big influence, deviations between data and NLO calculations 
are observed. This would indicate that higher order corrections are important.
It should however be noted that the dominant theoretical uncertainty, coming
from variations in the renormalisation and factorisation scales, completely
account for the observed deviations. Thus further constraints of the density
functions require a better theoretical understanding of the scale dependence.   

\section{Di-jet production associated with charm}

As has already been observed in Fig.\ref{tridiff}, measurements have clearly 
shown that the influence of the partonic structure 
of the photon decreases as its virtuality increases. The ZEUS collaboration
has studied the cross section ratio of direct and resolved photon processes
for di-jet events containing a $D^\star$-meson. Data have 
been compared 
to predictions of the HERWIG Monte Carlo program using two options of
the SAS1D parametrisation for the virtual photon. One option considers a
suppression of the partonic structure with increased photon virtuality,
whereas the other does not. 

\begin{figure}
\vskip 0.1cm
\begin{center}
\epsfig{figure=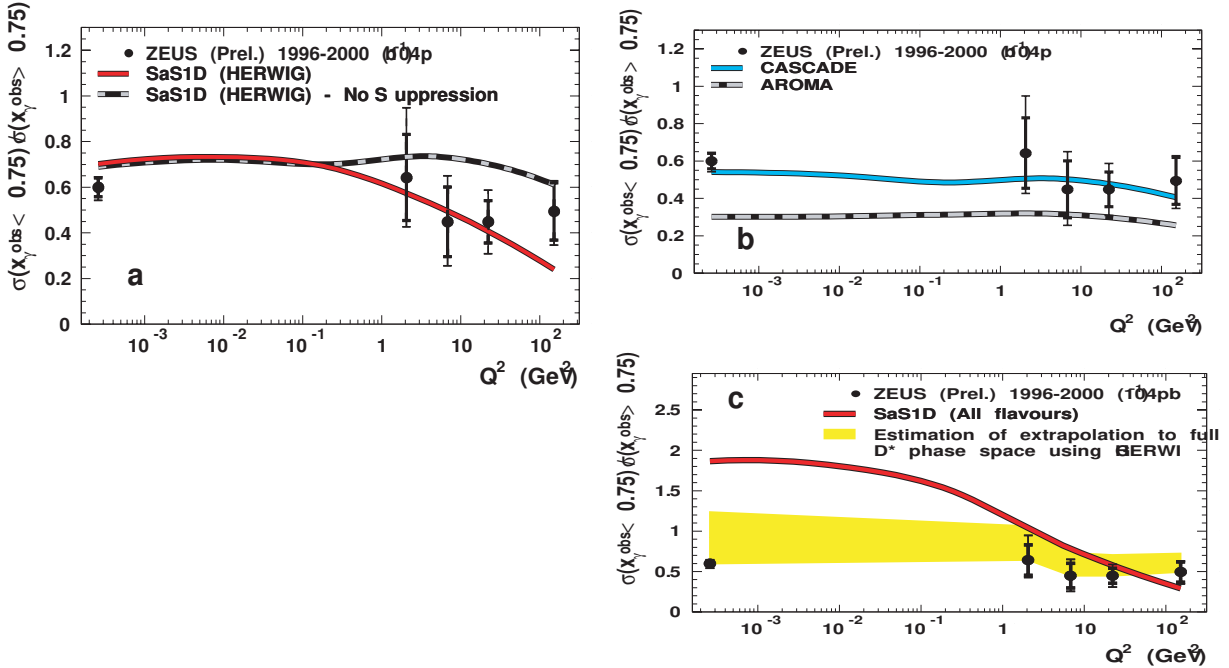,height=3.5in}
\end{center}
\caption{Ratio of low to high $x_\gamma$ for di-jet events with a
$D^\star$ meson. Data are compared to (a) the predictions of the SAS1D 
photon structure function, (b) the AROMA and CASCADE Monte Carlo 
programs and (c) the SAS1D photon structure function without the
requirement of a $D^\star$.
\label{zeusf3}}
\end{figure}

From Fig.\ref{zeusf3}a it is evident that the 
present data
do not have the necessary precision to be sensitive to a possible 
$Q^2$-suppression as given by SAS1D. However, the absence of an overall 
$Q^2$-dependence of the cross section ratio in this data
stays in sharp contrast to what has been observed for di-jet events
without any charm requirement. This fact indicates that the suppression 
at low $x_\gamma$ due to the charm requirement and due to the virtuality
are not independent.
Fig.\ref{zeusf3}b gives a comparison of data with two
models, which do not include any partonic structure of the photon. The AROMA 
model \cite{aroma} is based on the conventional DGLAP evolution scheme, whereas the
CASCADE model uses the CCFM evolution equations. The data clearly favour 
the CCFM based description with non $k_t$-ordered parton emissions in the
initial state. The ratio of di-jet events containing a $D^\star$-meson in one of the
jets is shown in Fig.\ref{zeusf3}c and compared to what is expected from
the description of SAS1D including all flavours but not requiring any 
$D^\star$-meson. The SAS1D predicion shows a clear $Q^2$-suppression
inconsistent with data. However, it was found that the kinematic cuts 
applied to select $D^\star$-mesons cause a suppression of the resolved
contribution. The HERWIG Monte Carlo program was used to extrapolate
the cross section ratio to the full phase space for $D^\star$-production.
This is given by the shaded band in the figure and is consistent with 
no $Q^2$-suppression in contrast to what is expected if no charm is
required.

Photoproduction of charm is an ideal testing ground for studying the 
underlying parton dynamics, since charm is predominantly produced by
$\gamma \rightarrow c \bar{c}$. The observation of an emission (jet)
with $p_t>p_t^c(p_t^{\bar{c}})$ indicates a scenario which is possible
in DGLAP only in a full $O(\alpha_S^2)$ calculation or
when charm excitation of the photon is included (the resolved
photon case). However, in the BFKL description such a scenario
comes naturally, since the transverse momenta along the 
evolution chain are not ordered.
The measured $x_\gamma$ 
spectrum  shows a tail to
small values, indicating that the hardest emission is
indeed not always coming from the charm quarks. Simulations with the CASCADE
generator shows that a significant part of the cross section
comes from events, where the gluon is the jet with the largest transverse
momentum. 
 A comparison of the measurement 
from the ZEUS collaboration with
the prediction from the full event simulation of
CASCADE using the unintegrated gluon distribution as given by 
Jung and Salam \cite{js} and applying 
jet reconstruction and jet selection at the hadron level shows
reasonably good agreement \cite{bjjpz}.

\section{Conclusions}

Studies of the real and virtual photon structure have clearly shown that 
NLO calculation and Monte Carlo generators using the DGLAP evolution
scheme are able to give reasonable agreement with data only if contributions
from resolved photon processes are included. On the other hand, the CASCADE
Monte Carlo program, in which the CCFM evolution equations are implemented,
gives an equally good description of the data but with less free parameters.  

\section*{Acknowledgments}
I have profited from helpful discussion with M. Derrick, E. Elsen, H. Jung 
and P. Schleper. The Swedish Research Council is acknowledged for financial
support.

\end{document}